# Cross Correlation-based Direct Positioning for Wideband Sources using Phased Arrays

Joon Wayn Cheong, and Andrew G Dempster, *Senior Member, IEEE*

*Abstract*— Recent developments in Phased Array direct positioning methods have improved accuracy for passively geo-locating multiple radio frequency-emitting signal sources. However, the number of geo-localisable signal sources is still limited by the number of antenna elements at each node. This is the limitation for methods based on MUSIC, otherwise known as signal subspace identification.

This paper attempts to exploit properties of wideband signal sources to compartmentalise signals into their respective Time Differences of Arrival. By performing direct positioning after the compartmentalisation process, we will show that geolocation of a large number of sources can be achieved by our proposed method at accuracies that exceed all existing methods, especially under low signal-to-noise ratio conditions.

## I. Introduction

It is well known that phased arrays can be used to passively determine the Direction of Arrival (DOA) of a Radiofrequency (RF) emitting signal source [1]–[3]. By deploying multiple stations of phased arrays that are geographically dispersed, DOA estimates retrieved from each station can be used to accurately geolocate a source using techniques such as least-squares that is based on the intersection of DOA lines of position [4]. The geo-localisation process becomes more complicated when there is more than one source as there are undesired intersections between DOAs from different sources.

If the source signals are sufficiently wideband, then cross-correlating signals received from two sites can determine the Time Difference of Arrivals (TDOAs) from these sources [3]. Like DOA-based geo-localisation, TDOA-based geo-localisation suffers from the same problem; that is, there are geo-localisation ambiguities arising from intersections of TDOAs from different sources.

The above approaches are sometimes called two-step or conventional algorithms, whereby a set of parameters (i.e. DOA or TDOA) is first determined from the signal stream, and then, the parameters are passed to a geo-localisation process to triangulate or trilaterate the estimated locations of the sources.

Direct positioning (DP) solves the multi-source geo-localisation problem via a Maximum-Likelihood (ML) approach [5]–[11]. The ML concept has also benefited Global Navigation Satellite System (GNSS) localisation in weak signal environment [12]–[17]. By directly manipulating digitised signals from all antenna elements from all stations, DP evaluates a predefined grid of position hypotheses to estimate the source positions. This is sometimes dubbed the one-step algorithm. Initial algorithms (such as Direct Position Determination) (DPD) based on this concept yielded superior accuracies than two-step algorithms [6], [9]. Further work derived from it enhanced separability between sources and accuracy by exploiting various instances of space-time relationships [18]–[20].

In DPD [6], geo-localisation of sources is first done for wideband sources by representing each source signal as a sum of multiple narrowband signals. These algorithms process the received signals as independent frequency channels and finally non-coherently combine the output of each frequency channel to geolocate the sources. The frequency splitting and noncoherent recombining process erodes the time delay information contained within the source's emitted wideband signal. Based on the same frequency decomposition principle, LOST [20]–[23] was derived to coherently combine signals from multiple sources to yield higher accuracy. However, this comes at the expense of a significant increase in computational resources especially for scenarios with large time-bandwidth product. Subsequently, LOST-FIND [19] was derived to cope with varied source signal bandwidth conditions.

Neither DPD nor LOST attempts to utilise the cross-correlation properties of wideband signals exploited by conventional TDOA estimators. TARGET [24] is the first DP algorithm that attempts to utilise these cross-correlation properties for geo-localisation. TARGET has demonstrated better source separability (i.e. it is able to discriminate between closely separated sources) in addition to enhanced geo-localisation accuracy. TARGET has geo-localisation limitations with regards to its detectable number of sources. This is linked to the limited number of antenna elements. A

This paragraph of the first footnote will contain the date on which you submitted your paper for review. It will also contain support information, including sponsor and financial support acknowledgment. For example, "This work was supported in part by the U.S. Department of Commerce under Grant BS123456".

This work was supported by the Australian Research Council (ARC) Linkage Funding LP140100252. This work is also supported by GPSat Systems Australia Pty Ltd with inputs from Ryan Thompson, Greame Hooper and Joe Fleming.

The authors are with the Australian Center for Space Engineering Research and the School of Electrical Engineering and Telecommunications, University of New South Wales, Sydney NSW 2052, Australia. (e-mail: cjwayn@unsw.edu.au).
.

recent paper [25] has attempted to address TARGET's limitation to the number of detectable sources.

This paper concentrates on DP methods stemming from MUSIC (e.g. [6], [8], [9], [20]) which in recent years has been merged with cross correlation-based methods such as TARGET to have the highest accuracy and separability in geo-localising sources given no *a priori* knowledge of the source's signal.

DP methods stemming from MUSIC involve three steps: (i) covariance matrix evaluation, (ii) signal subspace identification and (iii) geo-location via cost function minimisation.

The main contributions of this paper are our close examination of the working principle of TARGET (see Section IV) and our enhancement to it to derive a method that compartmentalises the signals based on their TDOAs before signal subspace identification. As justified in section V, the derived method will fully exploit both AOA and TDOA information embedded in the received signal streams for geo-localising multiple sources, performed at the cost function minimisation step. Our numerical results further show the superior performance of our proposed method when compared against DPD, TARGET and LOST in underdetermined scenarios. Its significance is especially pronounced at low signal-to-noise ratio (SNR) conditions.

## II. Signal Model

Consider $Q$ RF sources transmitting wideband signals of unknown waveform that can be observed by L sensor stations. Each sensor station is equipped with an M-element phased array. The M x 1 vector of complex baseband signals observed at the $l$-th station array can be modelled as:

$$\boldsymbol{r}_l(t) = \sum_{q=1}^{Q} \alpha_{l,q} \boldsymbol{a}_l(\boldsymbol{p}_q) s_q\left(t - \tau_{l,q}(\boldsymbol{p}_q)\right) + \boldsymbol{n}_l(t) \\ = \boldsymbol{A}_l \boldsymbol{\Omega}_l \boldsymbol{s}_l(t) + \boldsymbol{n}_l(t) \quad (1)$$

where $\alpha_{l,q}$ is the complex scalar representing the path attenuation coefficient for sensor $l$ and source $q$, $\boldsymbol{a}_{l,q}(\boldsymbol{p})$ is the M x 1 vector of complex coefficients representing the antenna's steering vector in response to the source at location $\boldsymbol{p}_q$, $s\left(t - \tau_l(\boldsymbol{p}_q)\right)$ is the complex baseband representation of the transmitted source signal waveform delayed by $\tau_l(\boldsymbol{p}_q)$ and $\boldsymbol{n}_l(t)$ is the M x 1 vector of complex Gaussian white noise. Thus, the matrix representations for the AOA-dependant antenna steering vectors, source signal and path attenuation coefficients are respectively:

$$\boldsymbol{A}_l = [\boldsymbol{a}_l(\boldsymbol{p}_1) \dots \boldsymbol{a}_l(\boldsymbol{p}_Q)]$$
$$\boldsymbol{s}_l(t) = [s_1\left(t - \tau_{l,1}(\boldsymbol{p}_1)\right) \dots s_Q\left(t - \tau_{l,Q}(\boldsymbol{p}_Q)\right)]$$
$$\boldsymbol{\Omega}_l = (\delta_{j,q} \alpha_{l,q})_{1 \leq q \leq Q}$$

### A. Narrowband Decomposition of Wideband Signals

As described in classical AOA systems, a narrowband assumption is valid when the time bandwidth product satisfies

$$\Delta \tau_{max} B \ll 1 \quad (2)$$

where $\Delta \tau_{\max} = \max_q \max_i \max_j |\tau_{j,q}(\boldsymbol{p}_q) - \tau_{i,q}(\boldsymbol{p}_q)|$ is the largest possible propagation time across the network of stations. $\tau_{j,q}(\boldsymbol{p}_q)$ and $\tau_{i,q}(\boldsymbol{p}_q)$ is the propagation time between source $q$ and station $j$ and station $i$ respectively. $B$ is the bandwidth of the source signal.

If the condition holds, then the narrowband assumption satisfies the approximation $s_q(t - \tau_l) \approx s_q(t) e^{-j 2\pi f_o \tau_l}$, where $s_q(t) \in \mathbb{C}$ is the signal of source $q$ and $f_o$ is the center frequency. In cases where the narrowband assumption fails, a frequency channelization method of decomposition can be adopted such that an individual frequency channel has bandwidth $B'$ that would satisfy the time bandwidth product $\Delta \tau_{max} B' \ll 1$ for each frequency channel. Therefore, the signal model for a wideband transmitter can be frequency decomposed into frequency subchannels $\bar{s}(k, t)$:

$$s_q\left(t - \tau_{l,q}(\boldsymbol{p}_q)\right) \approx \sum_{k=1}^{K} \bar{s}_q(k, t) e^{-j 2\pi (f_o + \Delta f_k) \tau_{l,q}} \quad (3)$$

DPD [6], [9] is one method that performs this form of narrowband decomposition to cope with wideband signals. In the case of DPD, K denotes the number of frequency channels, which is also lower bounded by the time-bandwidth product [6], [9].

### B. Cross-correlating Wideband Signals

In this paper, we consider multiple sources emitting wideband signals which satisfy a sufficiently large time-bandwidth product:

$$\Delta \tau_{\max} B \gg 1 \quad (4)$$

where the product of the largest propagation time across the sensor network $\Delta \tau_{\max}$ and the source signal bandwidth $B$ is sufficiently large to satisfy (4).

A source signal that is sufficiently wideband will exhibit a distinct amplitude peak in the cross-correlation function between the signals received at two stations with sufficiently wide geographical separation. Expressing this for the case of two antenna arrays and choosing the simplistic case where there is only one source as an example, cross-correlating signals from antenna arrays from widely-spaced stations $i$ and $j$ yields:

$$E\{\boldsymbol{r}_i(t) \boldsymbol{r}_j^H(t - \tau)\} \approx \begin{cases} \boldsymbol{a}_i \boldsymbol{\Omega}_i \bar{s}(\tau) \boldsymbol{\Omega}_j^H \boldsymbol{a}_j^H & , \tau \approx \Delta \tau_{ij} \\ 0 & , \tau \neq \Delta \tau_{ij} \end{cases} \quad (5)$$

Here, we have used the shorthand $\Delta \tau_{ij} = \tau_j - \tau_i$ as the true source TDOA between stations $i$ and $j$. $\bar{s}(\tau)$ is the auto-correlation of the signal at time delay $\tau$. For brevity noise terms has been dropped in (1).

This is the property that TARGET and our proposed method (i.e. ccDPD) will exploit. For these methods, the resolution of the TDOA grid is at least $1/B$ and the number of TDOA grid points K is at least $\Delta\tau_{max} B$ [20], [24]. Hence, the number of frequency channels to be used for narrowband decomposition and the number of TDOA grid points used for cross-correlation techniques are the same, (i.e. $K$).

### III. DIRECT POSITIONING

Traditionally, geo-localising an interference source is done for wideband sources by splitting the signal into multiple narrowband signals. This section examines the DPD and LOST geo-localisation methods.

*A. Geo-localisation using DPD*

We can concatenate $r_l(t)$ to form the composite signal matrix $y(t) = [r_1^T(t) \quad ... \quad r_L^T(t)]^T$. Its corresponding covariance matrix would then be $R_y = E_t\{y(t)y^H(t)\}$.

Considering a nominal scenario and given known number of transmitters $Q$, the eigen-decomposition of the covariance signal would produce the signal and noise subspaces denoted by $\Psi_y$ and $\Phi_y$, respectively. The covariance matrix $R_y$ can then be represented by the following

$$R_y = [\Psi_y \quad \Phi_y]\begin{bmatrix} \Sigma_y^2 + \sigma_y^2 I & 0 \\ 0 & \sigma_y^2 I \end{bmatrix}\begin{bmatrix} \Psi_y^H \\ \Phi_y^H \end{bmatrix}$$

This step of signal subspace identification can be done via eigen-decomposition. The matrix of eigenvectors is

$$\Psi_y = [u_1 \quad ... \quad u_Q]$$

$$u_q \in span\left(\begin{bmatrix} a_1(p_1) \\ \vdots \\ a_L(p_1) \end{bmatrix}, ..., \begin{bmatrix} a_1(p_Q) \\ \vdots \\ a_L(p_Q) \end{bmatrix}\right)$$

Thus, the signal subspace and noise subspace are represented by $\Psi_y$ and $\Phi_y$ respectively. The noise subspace is sometimes known as the orthogonal subspace and will always satisfy $\Psi_y + \Phi_y = I$.

Then DPD seeks to maximize the following cost function [9]

$$\hat{p} = \underset{p}{\text{argmax}}\, Q_y(p) = \underset{p}{\text{argmax}}\, \lambda_{max}\left(D_y(p)\right)$$

where

$$D_y(p) = H^H\hat{A}^H(p)\Psi_y\Psi_y^H\hat{A}(p)H$$

$$\hat{A}(p) = diag\{a_1^T(p), ..., a_L^T(p)\}$$

$$H \triangleq I_L \otimes 1_M$$

$\lambda_{max}(Z)$ is the largest eigenvalue of $Z$. Note that the steering vectors $a_l(\cdot)$ in $\hat{A}(\cdot)$ are unit vectors and needs to satisfy $a_l^H(\cdot)a_l(\cdot) = 1$. $diag\{\cdot\}$ is the function that concatenates vectors into a block diagonal form.

To accommodate the wideband case, the DPD algorithm described by [6] would result in a cost function that accommodates K frequency channels $\tilde{y} = [y_1^T(t) \quad ... \quad y_K^T(t)]^T$ with $D_y(p)$ replaced by $D_{\tilde{y}}(p)$ and

$$D_{\tilde{y}}(p) = H^H \sum_k \left[\hat{A}_{\tilde{y}}^H(k,p)\Psi_{\tilde{y}}(k)\Psi_{\tilde{y}}^H(k)\hat{A}_{\tilde{y}}(k,p)\right] H$$

where $\Psi_{\tilde{y}}(k)$ is the signal subspace $\Psi_y$ for the k-th frequency channel and the frequency channel dependent antenna steering matrix is,

$$\hat{A}_{\tilde{y}}(k,p) = diag\{a_1^T(p)e^{-i\omega_k\tau_1(p)}, ..., a_L^T(p)e^{-i\omega_k\tau_L(p)}\}$$

Notice that the cost function $D_{\tilde{y}}(p)$ effectively sums the squared contributions from individual frequency components. This non-coherent integration method lacks any function that exploits the cross-correlation properties of wideband signals.

*B. Geo-localisation using LOST*

Another wideband geo-localization processing, LOST [20] exploits the space-time covariance matrix concatenated from multiple time delayed signal streams $y(t)$ to form the composite signal matrix

$$z(t) = [y^T(t) \quad y^T(t+T_s) \quad ... \quad y^T(t+(K-1)T_s)]^T$$

Its corresponding covariance matrix would then be $R_z = E_t\{z(t)z^H(t)\}$. $R_z$ can then be represented by the following via eigen-decomposition

$$R_z = [\Psi_z \quad \Phi_z]\begin{bmatrix} \Sigma_z^2 + \sigma_z^2 I & 0 \\ 0 & \sigma_z^2 I \end{bmatrix}\begin{bmatrix} \Psi_z^H \\ \Phi_z^H \end{bmatrix}$$

The signal subspace is defined as $\Psi_z \in \Re^{KLM \times KQ}$ and the orthogonal subspace $\Phi_z \in \Re^{KLM \times (KLM-KQ)}$. This process of subspace identification (i.e. eigen-decomposition) is highly resource intensive especially when the time-bandwidth product is large. The final step of cost function computation according to the following relationships can be used to describe LOST [20]

$$J_{LOST}(p,\rho,f) = \frac{b^H(p,\rho,f)\Phi_z b(p,\rho,f)}{b^H(p,\rho,f)b(p,\rho,f)}$$

where,

$$b(p,\rho,f) = V(p)\alpha(\rho,f)$$

$$V(p) = I_K \otimes \hat{A}(p)$$

$$\Omega(f) = [1 \quad ... \quad (e^{j2\pi fT_s})^{K-1}]^T$$

$$\alpha(\rho,f) = \Omega(f) \otimes \left(diag\left(e^{j2\pi f\tau_1(p)}, ..., e^{j2\pi f\tau_L(p)}\right)\rho\right)$$

The path coefficients and frequency are denoted as $\rho \in \mathbb{C}^{LM}$ and $f \in \mathbb{R}_+$, respectively.

Consequently, the problem of seeking the minimum of $J_{LOST}(\boldsymbol{p},\boldsymbol{\rho},f)$ can be reduced to the following,

$$J_{LOST}(\boldsymbol{p}) = \lambda_{min}\left\{\left(\boldsymbol{V}^H(\boldsymbol{p})\boldsymbol{V}(\boldsymbol{p})\right)^{-1}\boldsymbol{V}^H(\boldsymbol{p})\boldsymbol{\Phi}_z\boldsymbol{V}(\boldsymbol{p})\right\}$$

Under nominal cases, where $\boldsymbol{p}$ approaches $\boldsymbol{p}_q$, the true location of $q$-th source, it also satisfies $\boldsymbol{a}_l(\boldsymbol{p}) \approx \boldsymbol{a}_l(\boldsymbol{p}_q), \forall l \in [1, L]$, which is where some local minimum for $J_{LOST}(\boldsymbol{p})$ is obtained as it approaches the following asymptote.

$$\lim_{\boldsymbol{p}\to\boldsymbol{p}_q} \boldsymbol{V}^H(\boldsymbol{p})\boldsymbol{\Phi}_z\boldsymbol{V}(\boldsymbol{p}) = \boldsymbol{0}$$

As $\boldsymbol{\Phi}_z$ is a subspace that encompasses all frequency channels, LOST coherently integrates across the channels. This leads to superior performance for a small number of sources. However, note that the subspace $\boldsymbol{\Phi}_z$ contains signal contributions from all sources, which we will contrast in the next section.

## IV. CROSS CORRELATION-BASED SOURCE LOCALISATION

DPD attempts to address wideband source geo-localisation by non-coherently combining the various frequency components, thus eliminating the TDOA information embedded within the signals $s(t - \tau_l(\boldsymbol{p}))$. A method that doesn't employ frequency splitting is TARGET [24]. TARGET uses cross-correlation between signals from all nodes as its input. For completeness, we briefly describe the TARGET algorithm here to draw comparison against our proposed method in the next section.

### A. TARGET

In fact, it is possible to discriminate and compartmentalise signals from different sources based on their expected TDOA. TARGET is one such method. We first concatenate two $\boldsymbol{r}_l(t)$ vectors from station $i$ and $j$ to form the observed signal vector $\boldsymbol{x}_{i,j}(t,\tau)$

$$\boldsymbol{x}_{i,j}(t,\tau) = \begin{bmatrix}\boldsymbol{r}_i^T(t) & \boldsymbol{r}_j^T(t+\tau)\end{bmatrix}^T \quad (7)$$

Then, we can describe (7) based on (1) as

$$\boldsymbol{x}_{i,j}(t,\tau) = \sum_{q=1}^{Q}\boldsymbol{U}(\boldsymbol{p}_q)\boldsymbol{s}_q(t,\tau) + \boldsymbol{n}_l(t,\tau)$$

$$\boldsymbol{U}(p_q) = \begin{bmatrix}\alpha_{i,q}\boldsymbol{a}_i(\boldsymbol{p}_q) & 0 \\ 0 & \alpha_{j,q}\boldsymbol{a}_j(\boldsymbol{p}_q)\end{bmatrix}$$

$$\boldsymbol{s}_q(t,\tau) = \begin{bmatrix}s_q(t - \tau_{i,q}(\boldsymbol{p}_q)) \\ s_q(t - \tau_{j,q}(\boldsymbol{p}_q) + \tau)\end{bmatrix}$$

Observe that when $\tau$ is correctly aligned with the $q$-th source position, i.e. $\tau \approx \tau_{j,q}(\boldsymbol{p}_q) - \tau_{i,q}(\boldsymbol{p}_q) = \Delta\tau_{i,j}(\boldsymbol{p}_q)$, (7) becomes

$$\boldsymbol{x}_{i,j}\left(t,\Delta\tau_{i,j}(\boldsymbol{p}_q)\right) = \sum_{q=1}^{Q}\boldsymbol{u}(\boldsymbol{p}_q)s_q\left(t - \Delta\tau_{i,j}(\boldsymbol{p}_q)\right) + \boldsymbol{n}_l\left(t,\Delta\tau_{i,q}(\boldsymbol{p}_q)\right)$$

where $\boldsymbol{u}(\boldsymbol{p}_q) = \boldsymbol{U}(\boldsymbol{p}_q) \times \boldsymbol{1}_{2\times 1}$. In other words, when the TDOA hypothesis $\Delta\tau_{i,j}$ based on the hypothesised position $\boldsymbol{p}_q$ is correct, then the expected covariance matrix of $\boldsymbol{x}_{i,j}$ will turn into a rank one matrix as opposed to a rank two matrix.

The covariance matrix of $\boldsymbol{x}_{i,j}(t)$ expressed as $\boldsymbol{R}_x(\tau)$ can be shown as

$$\boldsymbol{R}_x(\tau) = \mathbb{E}_t\left(\boldsymbol{x}_{i,j}(t,\tau)\boldsymbol{x}_{i,j}^H(t,\tau)\right)$$

$$\boldsymbol{R}_x(\tau) = \begin{bmatrix}\boldsymbol{R}_{ii} & \boldsymbol{R}_{ij}(\tau) \\ \boldsymbol{R}_{ji}(-\tau) & \boldsymbol{R}_{jj}\end{bmatrix} + \boldsymbol{R}_n \quad (8)$$

where $\boldsymbol{R}_{ii} = \sum_q \boldsymbol{a}_i(\boldsymbol{p}_q)\Lambda_{ii}\boldsymbol{a}_i^H(\boldsymbol{p}_q)\alpha_{i,q}^2$ and $\boldsymbol{R}_{ij}(\tau) = \sum_q \alpha_{i,q}\boldsymbol{a}_i(\boldsymbol{p}_q)\Lambda_{ij}(\tau)\boldsymbol{a}_j^H(\boldsymbol{p}_q)\alpha_{j,q}^*$. Note that the cross-correlation matrix originates from $\boldsymbol{R}_{ij}(\tau) = \mathbb{E}_t\left(\boldsymbol{r}_i(t)\boldsymbol{r}_j^T(t+\tau)\right)$ and $\boldsymbol{R}_{ji}(\tau) = \boldsymbol{R}_{ij}^H(-\tau)$. The transmitted signals' inter-station autocorrelation and cross correlation functions are defined as $\Lambda_{ii} = \mathbb{E}_t\left(s_i(t)s_i^H(t)\right)$ and $\Lambda_{ij}(\tau) = \mathbb{E}_t\left(s_i(t)s_j^H(t+\tau)\right)$ respectively.

At incorrect position hypotheses, there will be minimal contribution from signal components to $\boldsymbol{R}_{ij}(\tau)$. Conversely, for a given $\tau$, $\boldsymbol{R}_{ij}(\tau)$ would not contain signal components from other sources that do not have $\tau$ as their true TDOA. Hence, unlike LOST and DPD, the covariance matrices $\boldsymbol{R}_x(\tau)$ and $\boldsymbol{R}_{ij}(\tau)$ are now dependant on the TDOA hypothesis (which could be implied by a position hypothesis). Therefore, $\boldsymbol{R}_x(\tau)$ and $\boldsymbol{R}_{ij}(\tau)$ are both a form of covariance matrix that would only accentuate signal components with TDOA corresponding to $\tau$, leading them to perform better in a multiple source scenario.

Although not stated explicitly in the original paper [22], [24], the following derivations are thought to be implied by the algorithm for position estimation. From [22], [24], we further define the matrix $\boldsymbol{G}_{ij}(\tau)$ as

$$\boldsymbol{G}_{ij}(\tau) = \boldsymbol{I}_M - \boldsymbol{R}_{ii}^+\boldsymbol{R}_{ij}(-\tau)\boldsymbol{R}_{jj}^+\boldsymbol{R}_{ji}(\tau) \quad (9)$$

where

$$\boldsymbol{R}_{ii}^+ = \boldsymbol{\Pi}_i\boldsymbol{\Sigma}^{-1}\boldsymbol{\Pi}_i^H \qquad \boldsymbol{R}_{ii} = \boldsymbol{\Pi}_i\boldsymbol{\Sigma}\boldsymbol{\Pi}_i^H$$

The $Q \times Q$ diagonal matrix $\boldsymbol{\Sigma}^{-1}$ is the inverse of $Q$ largest eigenvalues of $\boldsymbol{R}_{ii}$ and the corresponding $Q$ eigenvectors are represented by the $M \times Q$ matrix $\boldsymbol{\Pi}_i$.

Finally, the algorithm estimates the source position using the sum of Rayleigh quotients as

$$\widehat{p_q} = \underset{p \in P}{\mathrm{argmin}} \frac{1}{L(L-1)} \sum_{i \neq j} J_{T,ij}(p_q) \quad \forall i,j \in [1, L]$$

where the TARGET Rayleigh quotient is defined as:

$$J_{T,ij}(p_q) = \frac{a_i^H(p_q) G_{ij}(\tau(p_q)) a_i(p_q)}{a_i^H(p_q) a_i(p_q)}$$

While TARGET does exploit TDOA for geo-locating sources, it can be seen from (9) that performing eigen-decomposition for $R_{ii}^+$ and $R_{jj}^+$ limits the capability of the algorithm to cases where there is a limited number of sources where $Q < min_l M_l$. This is an obvious limitation that has been readily acknowledged by its authors [24]. It also fails to simultaneously exploit all TDOA pairs coherently for geo-localisation. Conventional cross-correlation methods [3] used by single-antenna multi-station systems using TDOA for geo-localisation do not have such strict limitations.

## V. A NEW CROSS CORRELATION-BASED ESTIMATOR

Based on the principles described by existing approaches, we derive a new method that circumvents the need to perform eigen-decomposition at the early stages of computation (the approach used by TARGET) and the need to perform narrowband decomposition (the approach used by DPD).

In the same spirit as (7), we first define the TDOA-compensated composite signal vector from all $L$ stations as:

$$y(t, \tau(\rho)) = \begin{bmatrix} r_1^T(t) & r_2^T(t + \tau_{1,2}) & \cdots & r_L^T(t + \tau_{1,L}) \end{bmatrix}^T$$

The vector of all TDOA pairs is denoted as $\tau(\rho) = [\tau_{1,2}, \tau_{1,3}, \cdots, \tau_{L-1,L}]^T$ where $\tau_{i,j} = \tau_j(\rho) - \tau_i(\rho)$. When $\rho = p_q$, its corresponding composite covariance matrix is

$$R_y(\tau) = \mathbb{E}_t(y(t, \tau) y^H(t, \tau))$$

$$R_y(\tau(\rho)) = \begin{bmatrix} R_{11} & R_{12}(\tau_{1,2}) & \cdots & R_{1L}(\tau_{1,L}) \\ R_{21}(-\tau_{1,2}) & R_{22} & \cdots & R_{2L}(\tau_{2,L}) \\ \vdots & \vdots & \ddots & \vdots \\ R_{L1}(-\tau_{1,L}) & R_{L2}(-\tau_{2,L}) & \cdots & R_{LL} \end{bmatrix} + R_n$$

Note that $R_y(\tau(\rho))$ reduces to a block diagonal matrix under an incorrect position hypothesis $\rho \neq p_q$, for $\forall q$ described as

$$R_y(\tau(\rho)) \approx \begin{bmatrix} R_{11} & 0_{M \times M} & \cdots & 0_{M \times M} \\ 0_{M \times M} & R_{22} & \cdots & 0_{M \times M} \\ \vdots & \vdots & \ddots & \vdots \\ 0_{M \times M} & 0_{M \times M} & \cdots & R_{LL} \end{bmatrix} + R_n \quad (10)$$

Under asymptotically high SNR, the covariance matrix $R_y(\tau(\rho))$ has rank $LQ$ under an incorrect position hypothesis $\rho \neq p_q$, whereas it reduces to rank $L(Q-1) + 1$ when $\rho = p_q$ is satisfied.

This characteristic of the composite covariance matrix $R_y(\tau(\rho))$ is undesirable as the number of sources Q will need to satisfy $Q < M$ for nominal operation of signal subspace, leading us to propose the following modified composite covariance matrix

$$\widetilde{R}_y(\tau(\rho)) = R_y(\tau(\rho)) - \overline{R}$$

where $\overline{R} = diag(R_{11}, \ldots, R_{LL})$ is a block diagonal matrix. Hence, $\widetilde{R}_y(\tau(\rho))$ is a matrix that is constructed only from $R_{ij}(\tau_{i,j})$ for $\forall i \neq j$.

$$\widetilde{R}_y(\tau(\rho)) = \begin{bmatrix} 0_{M \times M} & R_{12}(\tau_{1,2}) & \cdots & R_{1L}(\tau_{1,L}) \\ R_{21}(-\tau_{1,2}) & 0_{M \times M} & \cdots & R_{2L}(\tau_{2,L}) \\ \vdots & \vdots & \ddots & \vdots \\ R_{L1}(-\tau_{1,L}) & R_{L2}(-\tau_{2,L}) & \cdots & 0_{M \times M} \end{bmatrix} \quad (11)$$

The modified covariance matrix $\widetilde{R}_y(\tau(\rho))$ eliminates the autocorrelation matrices of each station with itself $R_{ii}$ which result in $\widetilde{R}_y(\tau(\rho))$ to be rank zero when $\tau(\rho) \neq \tau(p_q), \forall q$ under asymptotically high SNR. This then becomes a favourable feature to form a geo-localisation estimator. From this, we compute the orthogonal subspace of $\widetilde{R}_y(\tau(\rho))$ to obtain the cost function with respect to a position hypothesis $\rho$.

Observe that $\widetilde{R}_y(\tau(\rho))$ compartmentalises signals according to their respective TDOA hypotheses. Unlike TARGET, $\widetilde{R}_y(\tau(\rho))$ concatenates the cross-covariance matrix from all inter-station pairs. This structure allows $\widetilde{R}_y(\tau(\rho))$ to capture a larger dimension for the formation of a signal or orthogonal subspace, as will be seen next.

While the position hypothesis $\rho$ has dimension $D \in \{2,3\}$, a specific position hypothesis $\rho$ immediately implies a specific set of TDOA hypotheses for all $(L-1)L/2$ pairs of stations.

$$\tau(\rho) = [\tau_{1,2}, \tau_{1,3}, \ldots, \tau_{1,L}, \tau_{2,3}, \ldots \tau_{L-1,L}]$$

Unlike DPD, but similar to TARGET, this new approach calls for a different signal subspace for each position hypothesis. Hence, for a given full set of hypothesised TDOAs $\tau(\rho)$, or a given hypothesised position $\rho$, one of the following defined conditions will be met. There can be four defined conditions to characterise (11):

*condition 0:* TDOA hypotheses at all $L - 1$ station pairs for all Q sources are incorrect, i.e. $\tau_{1,l}(\rho) \neq \tau_{1,n}(p_q)$ for $\forall l \in \{2, \ldots, L\}$ and $\forall q \in \{1, \ldots, Q\}$.

*condition 1:* The TDOA hypotheses at all station pairs are correct for source q, i.e. $\tau(\rho) = \tau(p_q)$ for $\forall l \in \{2, \ldots, L\}$. No other sources share the same TDOA for all station pairs considered;

*condition 2:* This condition considers a partially correct set of TDOA hypotheses. TDOA hypotheses at $L - N - 1$ station pairs are correct for source q whereas the TDOA hypotheses at $N$ station pairs are wrong for source q. In other words, $\tau_{1,l}(\rho) = \tau_{1,l}(p_q)$ for some $l \in \{2,3, \ldots, L\} \backslash \{\forall n\}$, whereas $\tau_{1,n}(\rho) \neq \tau_{1,n}(p_q)$ for some $n \in \{2,3, \ldots, L\} \backslash \{\forall l\}$;

*condition 3:* TDOA hypotheses for $L-1$ station pairs are correct for source $q$ whereas some of the TDOA hypotheses also matches with the true TDOA of another source $\bar{\bar{q}}$ for $N$ stations ($N < L$), i.e. $\boldsymbol{\tau}(\boldsymbol{\rho}) = \boldsymbol{\tau}(\boldsymbol{p}_q)$ for $\forall l \in \{2, \ldots, L\}$ and at the same time $\tau_{1,l}(\boldsymbol{\rho}) = \tau_{1,l}(\boldsymbol{p}_{\bar{\bar{q}}})$ for some $l \in \{2, \ldots, L\}$. In other words, this is a TDOA hypothesis $\boldsymbol{\tau}(\boldsymbol{\rho})$ where source q satisfied *condition 1*, but at the same time another interfering source $\bar{\bar{q}}$ satisfied *condition 2*.

*Analysis of condition 0:*

The modified composite covariance matrix $\widetilde{\boldsymbol{R}}_y(\boldsymbol{\tau}(\boldsymbol{\rho}))$ approximates null rank under *condition 0* because all $\boldsymbol{R}_{ij}(\boldsymbol{\tau})$ matrices reduce to infinitesimally small values dominated by cross-correlation terms at the incorrect TDOA hypotheses. This way, the autocorrelation coefficients $\boldsymbol{R}_{ii}$ that appear in $\boldsymbol{R}_y(\boldsymbol{\tau}(\boldsymbol{\rho}))$ under *condition 0* have been eliminated in $\widetilde{\boldsymbol{R}}_y(\boldsymbol{\tau}(\boldsymbol{\rho}))$, ensuring that the signal component is minimised under the incorrect TDOA hypotheses.

*Analysis of condition 1:*

Let's consider *condition 1* for $q = 1$ (i.e. $\boldsymbol{\rho} = \boldsymbol{p}_1$) and $Q > 1$. The modified covariance matrix $\widetilde{\boldsymbol{R}}_y(\boldsymbol{\tau}(\boldsymbol{p}_1))$ would have rank $L$ and it would asymptote to the following under diminishing noise terms and minor $q \neq 1$ cross-correlation terms.

$$\widetilde{\boldsymbol{R}}_y(\boldsymbol{\tau}(\boldsymbol{p}_1)) \approx \begin{bmatrix} \boldsymbol{0}_{M \times M} & \cdots & \beta_{1L}\boldsymbol{a}_1(\boldsymbol{p}_1)\boldsymbol{a}_L^H(\boldsymbol{p}_1) \\ \beta_{12}^*\boldsymbol{a}_2(\boldsymbol{p}_1)\boldsymbol{a}_1^H(\boldsymbol{p}_1) & \cdots & \beta_{2L}\boldsymbol{a}_2(\boldsymbol{p}_1)\boldsymbol{a}_L^H(\boldsymbol{p}_1) \\ \vdots & \ddots & \vdots \\ \beta_{1L}^*\boldsymbol{a}_L(\boldsymbol{p}_1)\boldsymbol{a}_1^H(\boldsymbol{p}_1) & \cdots & \boldsymbol{0}_{M \times M} \end{bmatrix} \quad (12)$$

$\beta_{ij}$ are the complex coefficients that are the interactions of path coefficients arriving at i-th station and j-th station. The $L$ dominant eigenvectors $\boldsymbol{e}_{\widetilde{R}_y,n}$ of $\widetilde{\boldsymbol{R}}_y(\boldsymbol{\tau}(\boldsymbol{p}_1))$ would span the signal subspace $\boldsymbol{e}_{\widetilde{R}_y,n}(\boldsymbol{p}_1) \in S_q$. Of the L eigenvectors, one of which will correspond to the largest positive eigenvalue and $L-1$ of which will correspond to $L-1$ most negative eigenvalues. Under *condition 1*, the signal subspace of the $q$-th source is $S_q$:

$$S_q = \text{span}\left(\begin{bmatrix} \boldsymbol{a}_1(\boldsymbol{p}_q) \\ \boldsymbol{a}_2(\boldsymbol{p}_q) \\ \vdots \\ \boldsymbol{a}_L(\boldsymbol{p}_q) \end{bmatrix}, \begin{bmatrix} \boldsymbol{a}_1(\boldsymbol{p}_q) \\ \boldsymbol{0}_{M \times 1} \\ \vdots \\ \boldsymbol{0}_{M \times 1} \end{bmatrix}, \begin{bmatrix} \boldsymbol{0}_{M \times 1} \\ \boldsymbol{a}_2(\boldsymbol{p}_q) \\ \vdots \\ \boldsymbol{0}_{M \times 1} \end{bmatrix}, \ldots\right) \quad (13)$$

Hence, the eigenvector of the largest positive eigenvalue $\boldsymbol{e}_{\widetilde{R}_y,0}(\boldsymbol{p}_1)$ would be correlated with all $L$ antenna steering vector hypotheses given $\boldsymbol{\rho}$ for source $q = 1$.

*Analysis of condition 2:*

Let's consider *condition 2*, where only a partial set of the hypothesised TDOAs $\boldsymbol{\tau}(\boldsymbol{\rho})$ are correct for the $q$-th source at $L - N$ stations. Hence, the TDOA hypotheses are incorrect for N stations. Then, the eigen-decomposition of $\widetilde{\boldsymbol{R}}_y(\boldsymbol{\tau}(\boldsymbol{\rho}))$ will have only $L - N$ dominant eigenvectors that span only some but not all dimensions of the signal subspace. As an example, suppose that only stations 1 and 3 have the correct TDOA hypotheses for source $q$, then its signal subspace $S_q$ reduces to the following under *condition 2*:

$$S_q = \left\{\begin{bmatrix} \boldsymbol{a}_1(\boldsymbol{p}_q) \\ \boldsymbol{0}_{M \times 1} \\ \boldsymbol{a}_3(\boldsymbol{p}_q) \\ \vdots \end{bmatrix}, \begin{bmatrix} \boldsymbol{a}_1(\boldsymbol{p}_q) \\ \boldsymbol{0}_{M \times 1} \\ \boldsymbol{0}_{M \times 1} \\ \vdots \end{bmatrix}, \begin{bmatrix} \boldsymbol{0}_{M \times 1} \\ \boldsymbol{0}_{M \times 1} \\ \boldsymbol{a}_3(\boldsymbol{p}_q) \\ \vdots \end{bmatrix}\right\} \quad (14)$$

*Analysis of condition 3:*

*Condition 3* considers the case where source $q$ meets *condition 1* but coincides with the situation where source $\bar{\bar{q}}$ meets condition 2. Thus, the signal subspace of $\widetilde{\boldsymbol{R}}_y(\boldsymbol{\tau}(\boldsymbol{\rho}))$ will consist of contributions from both source $q$ and $\bar{\bar{q}}$.

As an example, suppose that L= 3, $\boldsymbol{\rho} = \boldsymbol{p}_1$, $q = 1$ and $\bar{\bar{q}} = 2$ and simultaneously the TDOA hypotheses $\boldsymbol{\tau}(\boldsymbol{\rho})$ satisfies $\tau_{1,3}(\boldsymbol{p}_q) = \tau_{1,3}(\boldsymbol{p}_{\bar{q}})$ and $\tau_{1,2}(\boldsymbol{p}_q) \neq \tau_{1,2}(\boldsymbol{p}_{\bar{q}})$. Then, the signal subspace of $\widetilde{\boldsymbol{R}}_y(\boldsymbol{\tau}(\boldsymbol{\rho}))$ denoted as $S_{q,\bar{q}}$ can be described as $S_{q,\bar{q}} \in \text{span}(S_q \cup S_{\bar{q}})$ where $S_q$ remains the same as the example in *condition 1* whereas $S_{\bar{q}}$ is $S_q$ as defined in *condition 2* in (10). Then, the eigen-decomposition of $\widetilde{\boldsymbol{R}}_y(\boldsymbol{\tau}(\boldsymbol{\rho}))$ under this condition will produce five dominant eigenvectors, two of which correspond to the two largest positive eigenvalues and three of which correspond to the three most negative eigenvalues.

Obviously, there can be more than one $\bar{\bar{q}}$ for *condition 3* and the above can be easily generalised for that case. However, the discussion here is restricted to one $\bar{\bar{q}}$ source for brevity.

From the above analysis, it is evident that the modified covariance matrix $\widetilde{\boldsymbol{R}}_y(\boldsymbol{\tau}(\boldsymbol{p}_1))$ benefits significantly by minimising interference issues caused by the occurrence of multiple sources. Recall *condition 1*, where the signal subspace formed is specific only to the set of TDOA hypotheses $\boldsymbol{\tau}(\boldsymbol{\rho})$. The signal subspace identification under *condition 3* is the only case where interference (a.k.a. multiple access) issues due to multiple source is encountered.

### A. Cross Correlation-based DPD

If there is only a single source and *condition 1* is achieved, the single most dominant eigenvector $\boldsymbol{e}_{\widetilde{R}_y,0}(\boldsymbol{p}_1)$ would then be used to form the orthogonal noise subspace $\boldsymbol{E}^{\perp}_{\widetilde{R}_y}(\boldsymbol{\rho})$ used in the Rayleigh cost function $J_{ccDPD}(\boldsymbol{\rho})$ in (16).

However, given the plausibility of multiple sources and *condition 3*, it becomes necessary to factor in more than one dominant eigenvector when identifying the signal subspace. To this end, we use the $N_m$ most dominant eigenvectors to reconstruct the approximated signal subspace given position hypothesis $\boldsymbol{\rho}$ as

$$\boldsymbol{E}_{\widetilde{R}_y}(\boldsymbol{\rho}) = \sum_{i=0}^{N_{max}-1} \boldsymbol{e}_{\widetilde{R}_y,i}(\boldsymbol{\rho})\boldsymbol{e}_{\widetilde{R}_y,i}^H(\boldsymbol{\rho})$$

Notice that $\boldsymbol{e}_{\widetilde{R}_y,i}(\boldsymbol{\rho})$ are the eigenvectors corresponding to $N_{max}$ largest eigenvalues of $\widetilde{\boldsymbol{R}}_y(\boldsymbol{\tau}(\boldsymbol{\rho}))$. $N_{max}$ is thus the maximum allowable TDOA collisions across the set of position grid $G$.

The corresponding Rayleigh cost function for $E_{\tilde{R}_y}(\rho)$ is thus,

$$J_{ccDPD}(\rho) = \lambda_{min}\left((A^H(\rho)A(\rho))^{-1}A^H(\rho)E_{\tilde{R}_y}^{\perp}(\rho)A(\rho)\right) \quad (15)$$

Where the noise subspace $E_{\tilde{R}_y}^{\perp}(\rho)$ and the antenna steering matrix $A(\rho)$ is

$$E_{\tilde{R}_y}^{\perp}(\rho) = I_{LM} - E_{\tilde{R}_y}(\rho)$$

$$A(\rho) = \begin{bmatrix} a_1(\rho) & \cdots & 0_{M\times 1} \\ \vdots & \ddots & \vdots \\ 0_{M\times 1} & \cdots & a_L(\rho) \end{bmatrix}_{LM\times L}$$

Note that $\lambda_{min}(Z)$ denote the smallest eigenvalue of $Z$. We can ease the computation of (15) via

$$J_{ccDPD}(\rho) = \frac{det\left(A^H(\rho)E_{\tilde{R}_y}^{\perp}(\rho)A(\rho)\right)}{det(A^H(\rho)A(\rho))} \quad (16)$$

Finally, geo-localisation is achieved by searching the grid of position hypotheses $G$ that minimises the ccDPD Rayleigh quotient:

$$\hat{\rho} = \underset{\rho \in G}{\arg\min} \; J_{ccDPD}(\rho)$$

From (16) and the description of the TDOA-dependant $\tilde{R}_y(\tau(\rho))$, we can see that ccDPD proposed here benefits simultaneously from both the processing gain of phased array beamforming and the processing gain of cross-correlating wideband signals, respectively.

## VI. NUMERICAL EVALUATION

To demonstrate the usefulness of the proposed method against several of the existing methods, an underdetermined scenario is chosen. That is, the number of sources $Q$ in the field exceeds the smallest number of antenna elements per array $M$. There are $L = 3$ base stations deployed in an equilateral triangle with a radius of 300m. Each base station consists of an M=8 element circular phased array with radius of 1.5λ where λ is the center frequency wavelength. The centre frequency is chosen as 1.575GHz, which is the center frequency for the GPS L1 band. The phased array receivers sample at $f_s = 20$MHz unless otherwise stated and a signal duration of one millisecond is considered to evaluate the covariance matrices.

$Q = 12$ sources are scattered in a 3-by-4 grid with random offsets, each transmitting a complex wideband White Gaussian Noise (WGN) signal with a bandwidth of B = 10MHz. This is said to be an overdetermined scenario whereby the number of sources $Q$ exceeds the number of antenna elements per array $L$. The geometry of the base stations and the sources are illustrated in Figure 1.

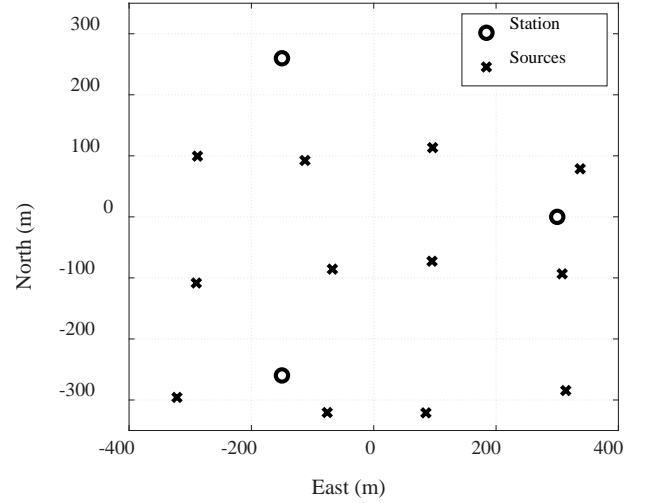

Fig. 1. Three base stations placed in an equilateral triangle with 12 sources placed at random offsets from a 3×4 grid.

We have implemented our proposed method (i.e. ccDPD) using parameter $N_m = 2$. A snapshot to visualise the cost function is shown in Figure 2. We have chosen SNR = −5dB for this example. Despite the challenging scenario setup, distinct cost function local minima can be seen at each source location without ambiguities

For comparison purposes, the existing DPD [6] and LOST [20] estimators are implemented using the parameter K = 35. The number of dominant eigenvectors extracted for LOST processing is chosen as LMQ = 288. The TARGET algorithm [24] extracts $Q_T = M - 1 = 7$ dominant eigenvectors for $R_{ii}$ signal subspace identification. A more recent method [26], denoted here as MChen2018, is also implemented using the same parameter as ccDPD, i.e. $N_m = 2$.

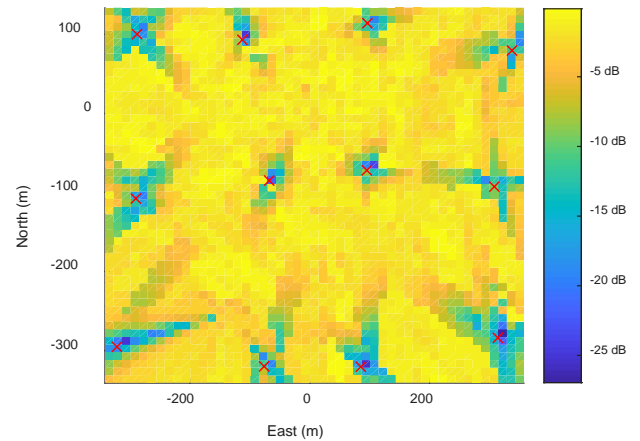

Fig. 2. Cost function of the proposed method, ccDPD at -5dB SNR. Red crosses indicate the true source locations.

The horizontal Root Mean Squared Error (RMSE) of each estimator is defined as $E = \sqrt{\frac{1}{QJ}\sum_{q=1}^{Q}\sum_{j=1}^{J}\|p_q - \widehat{p_{q,j}}\|^2}$ , where $J$ denotes the number of trials, $\widehat{p_{q,j}}$ is the estimated position of

$q$-th source at the $j$-th trial and $p_q$ is the true position of $q$-th source. The RMSE for the abovementioned methods for varying SNR conditions are shown in Figure 3.

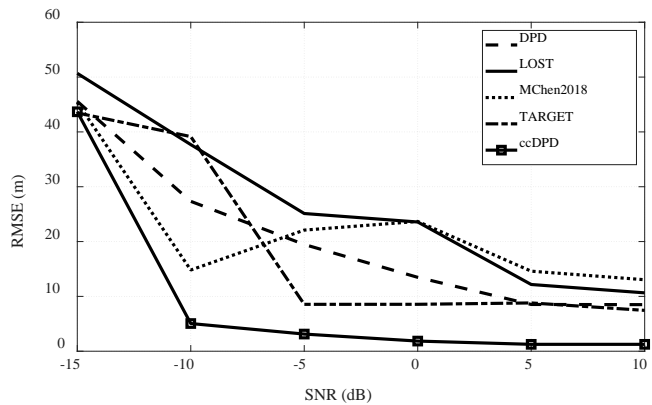

Fig. 3. Average RMS Error over various SNR for existing methods DPD [6], TARGET [24], LOST [20], MChen2018 [26] and our proposed method ccDPD.

It is evident from Figure 3 that ccDPD outperforms all other methods by a factor of 2 or better.

In addition to SNR variation, it is also important to compare ccDPD against existing methods under signal bandwidth variation. In this simulation, the frequency channelization parameter K for DPD is scaled down according to the relationship $K = \Delta\tau_{max} \times f_s$. Other estimator parameters are not varied as they are not bandwidth dependant. It is evident from Figure 4 that our proposed ccDPD method outperforms all existing methods except for bandwidths of 1MHz or smaller where DPD performs better.

TARGET is still a two-stage process: it first eigen-decomposes station-by-station before performing the second step of cost function evaluation, combination and estimation. This limits the dimension of the orthogonal subspace to be at most $L$. Because TARGET identifies the signal subspace station-by-station, it performs poorly at low SNR and the number of sources $Q$ are larger than the number of antenna elements $L$.

Our proposed ccDPD exploits TDOA compartmentalisation simultaneously from all stations. This produces a higher dimension covariance matrix with a higher dimension noise subspace than TARGET. In general, ccDPD has a similar order of magnitude in computational cost as TARGET whilst being more computationally advantageous than LOST.

## VII. CONCLUSION

Based on the analysis of existing Direct Positioning methods, we have proposed a way of manipulating the signal to produce a global covariance matrix that is TDOA dependant. Such a method effectively focusses the cost function to evaluate signal contributions pertaining only a specific set of implied TDOA hypotheses, i.e. minimising multiple access effects.

Our method produces more accurate geolocation results than all existing wideband methods in overdetermined scenarios under a wide range of Signal to Noise Ratios. In comparison to existing methods, a range of large bandwidth scenarios were also simulated to show ccDPD is indeed superior.

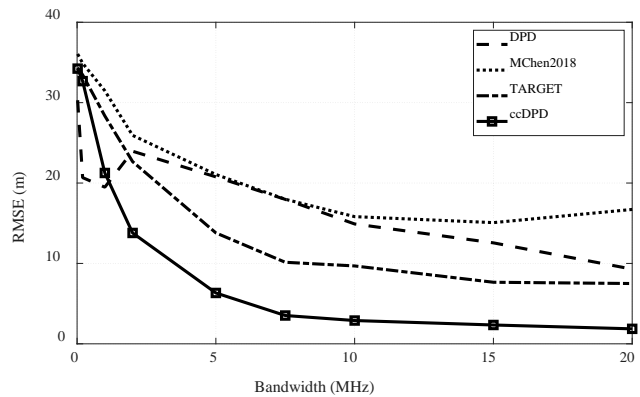

Fig. 4 Average RMS Error over various source signal bandwidths at a fixed SNR of -5dB for existing methods DPD [6], TARGET [24], MChen2018 [26] and our proposed method ccDPD.

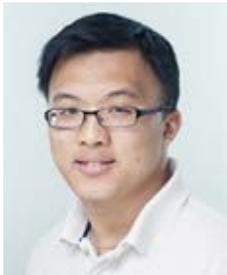

**Joon Wayn Cheong** is currently a Research Associate at the School of Electrical Engineering, University of New South Wales (UNSW) where he is currently developing the firmware for the space-qualified Namuru family of GPS/GALILEO integrated receivers under the Garada and QB50 project. He received his PhD from UNSW where he cracked the Locata pseudolite positioning system's code and derived high-sensitivity GPS signal acquisition algorithms. His other research interests in the GNSS field include weak signal acquisition, A-GPS, GNSS/pseudolite integrated signal processing, GNSS interference and spoofing.

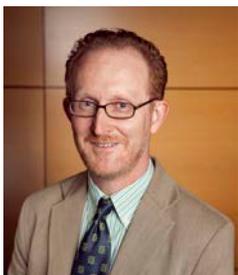

**Andrew G. Dempster** received the B.Eng. and M.Eng.Sc. degrees from the University of New South Wales (UNSW), Sydney, NSW, Australia, in 1984 and 1992, respectively, and the Ph.D. degree from the University of Cambridge, Cambridge, U.K., in 1995, in efficient circuits for signal processing arithmetic.

He is currently a Professor with the School of Electrical Engineering and Telecommunications at UNSW and the Director of the Australian Centre for Space Engineering Research (ACSER). He was a System Engineer and Project Manager for the first global positioning system receiver developed in Australia in the late 1980s and has been involved in satellite navigation ever since. He has published in the areas of arithmetic circuits, signal processing, biomedical image processing, satellite navigation, and space systems. His current research interests include satellite navigation receiver design and signal processing and space systems.